\documentclass[%
 reprint,
superscriptaddress,
nofootinbib,
 amsmath,amssymb,
 aps,
 prb,
]{revtex4-2}

\usepackage{graphicx}
\usepackage{dcolumn}
\usepackage{bm}

\usepackage[colorlinks = true,
linkcolor = black,
urlcolor  = blue,
citecolor = black,
anchorcolor = black]{hyperref}
\usepackage{longtable}

\usepackage{chngcntr}

\begin{document}


\title{A Rate Model of Electron Populations for Non-linear High-Fluence X-ray  Absorption Near-Edge Spectra}

\author{Robin Y. Engel}
    \affiliation{Deutsches Elektronen-Synchrotron DESY, Notkestr. 85, 22607 Hamburg, Germany}%
    \affiliation{Physics Department, Universität Hamburg, Luruper Chaussee 149, 22761 Hamburg, Germany}
\author{Markus Scholz}
    \affiliation{Deutsches Elektronen-Synchrotron DESY, Notkestr. 85, 22607 Hamburg, Germany}%
    \affiliation{European XFEL, Holzkoppel 4, 22869 Schenefeld, Germany}%
\author{Jan O. Schunck}
    \affiliation{Deutsches Elektronen-Synchrotron DESY, Notkestr. 85, 22607 Hamburg, Germany}%
    \affiliation{Physics Department, Universität Hamburg, Luruper Chaussee 149, 22761 Hamburg, Germany}
\author{Martin Beye}
    \affiliation{Deutsches Elektronen-Synchrotron DESY, Notkestr. 85, 22607 Hamburg, Germany}%
    \affiliation{Physics Department, Universität Hamburg, Luruper Chaussee 149, 22761 Hamburg, Germany}
\begin{abstract}
Absorbing a focused, femtosecond X-ray pulse from a Free-Electron Laser (FEL) can lead to extreme electronic excitations in solids. This excitation drives changes of the electronic system over the course of the pulse duration and the overall absorption of the pulse becomes fluence-dependent.
Thus, fluence-dependent non-linear X-ray Absorption Near Edge Spectroscopy (XANES) is sensitive to the valence excitation dynamics around the Fermi level on the few-femtosecond timescale.
Here we present a simplified rate model based on well-established physical mechanisms to describe the evolution of the electronic system. 

We construct temporal and spatial differentials for the processes of resonant absorption, stimulated emission, non-resonant absorption, Auger decay, valence band thermalization and scattering cascades of free electrons.

The phenomenological rate model approach provides a direct understanding how each physical process contributes to the fluence-dependent changes observed in XANES measurements.
Without accounting for fluence-dependent changes to the density of states, the model shows good agreement with experimental results on metallic nickel over more than three orders of magnitude in fluence,
establishing electron redistribution as the main driver of non-linear absorption changes at high fluences. 
Although in the closest vicinity of the resonance, more complex approaches are necessary to describe remaining discrepancies of the fluence-dependence changes,
the demonstrated capability to describe spectral changes up to extreme fluences yields fundamental insights into the complex dynamics after intense core excitation and provides an important tool for the design and evaluation of future FEL experiments, in particular for the development of non-linear X-ray spectroscopy. 
\end{abstract}

\maketitle


\section{\label{sec: introduction} Introduction}
Translating non-linear spectroscopy methods that are well established in the longer wavelength ranges \cite{boyd2020nonlinear, Demtroder2015NL_spect, Levenson1987nonlinearOptSpect} to the X-ray regime is particularly attractive, since the strong localization of transitions involving core electrons as well as element-specific absorption edges promise additional selectivity to these already potent analytical tools \cite{Beye2019musix, mukamel2005multiple, Glover2012xray, beye2013stimulated, weninger2013stimulated, Shwartz2014XSHG, Tamasaku2014TPA_competing, Bencivenga2015FWMTG, schreck2015implications, Lam2018HXSHGinterfacial, Tamasaku2018NL_TPA, higley2019femtosecond, rottke2021probing, Rouxel2021HXTG, Bencivenga2021FWMtwo_color, Wirok_Stoletow2022TPA_Ge, Boemer2021XoptWMprobes, Serrat2021locFWM, Schwickert2022Qcoherence}. 
A prerequisite for non-linear X-ray spectroscopy is a high density of X-ray photons, such as those found in the tightly focused X-ray radiation from a Free-Electron Laser (FEL). 
The absorption of such pulses drastically modifies the electronic structure of any investigated material, even on the timescale of femtosecond-short pulses. Absorption around material resonances directly probes transitions between the core-levels and states around the Fermi level and is thus sensitive to the electron dynamics in the valence band of materials.
Electronic structure changes in this region affect the degree of absorption experienced by later parts of the same pulse and the fluence-dependent absorption can be used to derive information on the excited state and its evolution  \cite{bob2009turning, vinko2010electronic, cicco2014interplay, yoneda2014saturable, rackstraw2015saturable, principi2016free,hantschmann2022rate, higley2019femtosecond, higley2022stimulated}.

Several models have been put forward to describe aspects of this interplay between photon absorption and electronic system \cite{hatada2014modeling, cicco2014interplay, hantschmann2022rate, higley2019femtosecond, higley2022stimulated}.
Away from material resonances, fluence-dependent X-ray absorption has been successfully modeled using rate models in three-level systems \cite{hatada2014modeling, cicco2014interplay} modeling a ground, core-excited and intermediate valence excited state. 
When probing the valence bands around material resonances however, three population parameters representing the state of the material at a given point in time become insufficient to represent the non-thermal electron energy distribution around the Fermi level in an extended solid that is relevant for the rates of excitation and relaxation.
This makes rate models using only three levels unsuitable to model non-linear absorption near resonances.

In this work, we present an expanded rate model to describe the evolution of the electronic system in terms of an energy-resolved population of the valence band within a constant Density of States (DoS). The valence electrons are heated through energy transfer from the scattering of free electrons from Auger decays and non-resonant absorption. 
The model uses material parameters known or calculated for the ground-state and scales these parameters in accordance with changes in electronic populations. 
Only the time constants for the valence band thermalization and the scattering cascades of free electrons are treated as free parameters.
We present calculations matching measurements of X-ray absorption spectra recorded with monochromatic X-rays in transmission through metallic nickel foils around the nickel 2$p_{3/2}$ ($L_3$) edge \cite{engel2022XFEL}; our model reproduces the main fluence-dependent changes in the measured spectra over more than three orders of magnitude. While the measurements are discussed in detail in a separate publication \cite{engel2022XFEL}, this paper illustrates the framework of the rate model and strives to provide an intuitive understanding of the mechanisms that drive non-linear changes.

The following text is structured as follows. 
In section (\ref{sec: model}), we give a qualitative overview of the rate model: Section \ref{sec: algorithm} describes the algorithm and approximations made in the interest of computational viability, Section \ref{sec: processes} outlines the mathematical formulation for the rates of all relevant processes, which are then assembled into differentials of the photon and electron populations in section \ref{sec: differentials}.
Section \ref{sec: parameters} elaborates on relevant input parameters.
Finally, we discuss the use and implications of our model in section \ref{sec: discussion}.

\section{\label{sec: model} The rate model}
We describe the propagation of X-ray photons through the sample as well as the dynamics of electron populations within the sample using a set of ordinary differential equations.
These rate equations describe the evolution of photon and electron densities and are assembled from terms that each describe a specific physical process. 
The rate of each process is based on ground-state rates, scaled with the appropriate fractional occupation (the number of electrons divided by the number of states). Each process rate is described in detail in section \ref{sec: processes}.
Such scaling inherently prevents any state from exceeding its physically meaningful population (between zero and the number of available states) and also enforces that the number of electrons in the sample is conserved over the simulated time.

The model allows for an arbitrary number of incident resonant photon energies $E_i$, for each of which a Gaussian temporal profile of incident intensity is assumed. 
For the presented calculations, only a single resonant photon energy was used, representing measurements with monochromatic X-rays. 
Incident photons are the only source of energy flow into the system, and all energy eventually contributes to the thermal energy of the valence system.

\begin{figure}[ht]
\centering
\includegraphics[width= \linewidth]{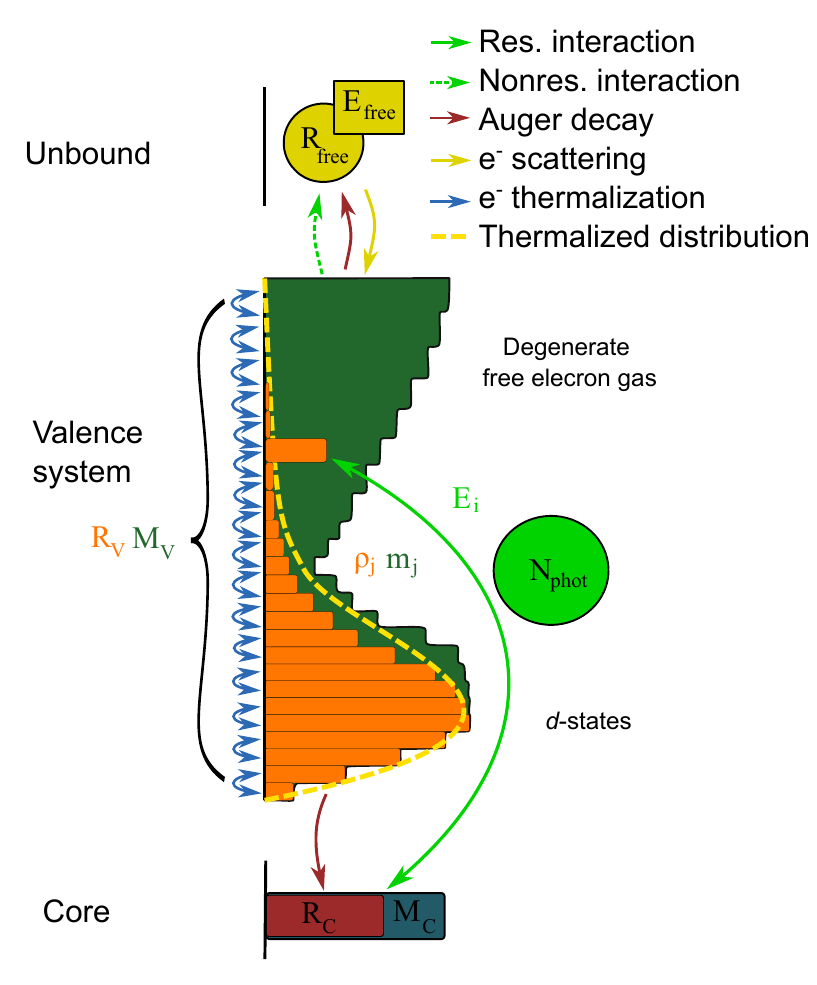}
\caption{
	\textbf{Photon-, electron- and energy-densities and their interactions.} A photon density $N_{phot}$ drives resonant interactions between the core electrons $R_C$ and specific valence electrons $\rho_j$. It also drives non-resonant excitations from the entire valence electron system $R_{V}=\sum_j \rho_j$ to free electrons $R_{free}$, which have a total energy of $E_{free}$. Auger decays transfer electrons from the valence system to both core states and free electrons; scattering cascades transfer electrons and energy from the free states to the valence system; thermalization drives the valence system towards a thermalized Fermi-Dirac distribution. $M_C, M_{V}$ and $m_j$ represent the number of available states and are pictured as bars to represent the energy bins of the numerical calculation.}
\label{fig: Variables}
\end{figure}

Figure \ref{fig: Variables} schematically summarizes the modeling of the electronic system: The electron populations $R_{C}$ and $R_{V}$ describe the total number of electrons bound in the core and valence system, respectively, for a single atom in the sample. Their values are limited by the number of available states, $M_{C}$ and $M_{V}$. 
In the presented example on the nickel L$_3$-edge spectra, the ground-state populations are $R_C = 4$, representing the 2p$_{3/2}$-electrons and $R_{V}=10$, representing electrons from the 3$d$ and 4$s$ states.
Electrons in intermediate shells (in the given example the 3$p$ and 3$s$ electrons) contribute to the non-resonant absorption but are not explicitly simulated.
We describe the electronic population of the valence system in an energy-resolved manner, splitting it up into a discrete number of densities $\rho_j$, where $j$ represents the index along the valence energy axis. 
The number of available states $m_j$ for each energy bin in the valence system is derived from the calculated ground-state DoS \cite{Persson2016nickel,Jain2013} up to 30\,eV above the Fermi level $E_F$. Beyond this value, the free electron gas model is used to calculate the density of states \cite{williams2020collisional}. We set the energy cutoff of the energy-resolved valence band at 800\,eV above the Fermi level. 
All electrons with even higher energies, such as photo-electrons created via non-resonant absorption and Auger-electrons from the decay of core-holes, are described in a separate pool of electrons $R_{free}$ without energy resolution, although the total energy in this pool is tracked by the parameter $E_{free}$.

Within our model, the DoS does not change over time since the DoS is dominated by the atomic lattice potentials and thus mainly affected by nuclear motion \cite{chen_two_temp}. 
Since the coupling of electronic excitations to nuclear motion is much slower than the purely electronic dynamics, observations that are temporally confined to the duration of a single FEL pulse (on the order of tens of femtoseconds) are usually dominated by electronic processes.

\subsection{Algorithm}
\label{sec: algorithm}

In contrast to a Monte-Carlo approach, where interaction pathways of many individual particles are computed and averaged, our rate model describes processes in terms of non-quantized, average quantities (densities) in a finite-element analysis.

The problem is formulated as an idealized X-ray pulse, Gaussian in time and space, traveling at the speed of light through a three-dimensional sample. The photon density interacts non-linearly with the sample. Each voxel of the sample is characterized by a complex electronic state as described above and in Figure \ref{fig: Variables}, which itself evolves in time after excitation.
To make this problem tractable, we apply additional key approximations that allow for a drastic reduction of the computational effort that would come with a naive implementation of this problem in a four-dimensional (three spatial dimensions and time) finite elements algorithm: First, we describe electronic excitations as strictly localized and neglect their propagation; instead, only photons propagate through the sample, and only in forward direction\footnote{We neglect spontaneous X-ray emission, which would constitute up to 0.9\,\% of nickel $L_3$ core-hole decays \cite{krause1979atomic}.}. 
Considering that a sample of 20\,nm thickness is traversed by light in less than 100\,as, the propagation of photons is calculated as if it happened instantaneous in between the time-steps performed to depict the evolution of the electronic system. This de-coupling of dimensional dependence effectively simplifies the problem into two sets of separately solvable, one-dimensional initial value problems: the photon propagation in space and the electronic evolution in time.

To solve these, we formulate the time-differential of all electronic states (see section \ref{sec: differentials}), depending on the photon density incident at a certain time.
The temporal evolution is solved using the fourth-order Runge-Kutta method with adaptive time-stepping based on a fifth-order local approximation \cite{2020SciPy-NMeth}. 
In the meantime, whenever a time-differential is computed, the current incident photon density is first propagated through the sample using the explicit fourth-order Runge-Kutta method (in space) to retrieve the photon density at each depth of the sample. 

This way, the time-dependent transmission for a Gaussian temporal pulse profile is calculated for a range of different overall pulse energies incident per sample area.
This treatment yet neglects the transversal profile of the beam. Because the absorption is non-linearly dependent on the incident fluence, the transmission of a transversally inhomogeneous beam must be integrated over a specific spot profile with spatially varying transmission.
Here, a two-dimensional Gaussian profile of the FEL spot is then accounted for by integrating the transmitted intensity based on previously calculated fluence-dependent transmission. The Figures \ref{fig: transmission}, \ref{fig: fermi solver}, \ref{fig: dos development} and \ref{fig: energy conservation} in the following section show intermediate results from simulations calculated for a specific fluence, as opposed to integrating over the transmission in a Gaussian spot-profile. 
With these simplifications, the overall computational complexity is drastically reduced.

\subsection{Processes}
\label{sec: processes}
The following describes the rates at which the physical processes occur for each atom. For a better overview, we introduce the processes as individual terms and assemble them into differential equations in the next section. In principle, each process is described using an absorption length or lifetime which is known from ground-state measurements and then scaled linearly with the changing electron populations with respect to the ground state. The normalization is such that the ground-state rate is reproduced for an undisturbed electron system and the rate vanishes when the corresponding transition cannot happen due to a lack of electrons or holes.
We use the indices $j$ and $i$ to refer to specific energies, where the index $i$ is used for photon energies of X-rays and the index $j$ for the energy of electronic states in the valence band.

\subsubsection{Resonant interaction}
The resonant interaction describes both resonant absorption (core-valence transitions) and stimulated emission (valence-core transitions) as a single process.
It is calculated for each energy $E_j$ in the valence system that is resonant with a given photon energy $E_i$. 

\begin{equation*}
    P^{res}_j = \left(\frac{R_C}{M_C} - \frac{\rho_j}{m_j} \right) \frac{N_i}{\lambda_i^{res}} \delta_{ij}
\end{equation*}
\begin{center}
\begin{tabular}{rl}
     $\lambda^{res}_i$:& Resonant absorption length \\
     $R_C$:& Number of core electrons   \\
     $M_C$:& Number of core states \\
     $\rho_j$:& Valence electrons at $E_j$\\
     $m_j$:& Valence states at $E_j$\\
     $N_i$:& Number of photons per $\mathrm{nm^2}$ at $E_i$\\
     $\delta_{ij}$:& Kronecker-delta\\
\end{tabular}
\end{center}

The first terms (in brackets) represent the difference in the occupation of core states $R_C/M_C$ and resonant valence states $\rho_j/m_j$. The dominance of absorption over stimulated emission or vice-versa is determined solely by this difference, as they represent an optically driven two-level system in the incoherent limit. 
If the core level population is smaller than the valence population, the resonant interaction process becomes negative, representing the dominance of stimulated emission.
The second term on the right is the number of irradiated photons divided by the penetration length. 
The Kronecker delta ensures that only photons and electrons in corresponding energy bins interact.

\subsubsection{Non-Resonant absorption}
The non-resonant absorption summarizes photon absorption from other electronic states than the resonant core-level, especially from the valence electrons. Photon densities $N_i$ at all incident energies reduce each population $\rho_j$.
\begin{equation*}
    P^\mathrm{non-res}_{i,j} = \frac{\rho_j}{R_{V}^0}  \frac{N_i}{\lambda^{non-res}}
\end{equation*}

\begin{center}
\begin{tabular}{rl}
     $\lambda^{non-res}$:& Non-resonant absorption length \\
     $R_{V}^0$:& Total number of valence electrons\\
                & in the ground state\\
\end{tabular}
\end{center}

The interaction is normalized by the total valence band population in the ground state $R_{V}^0$, so that the sum of the first term over all $j$ becomes unity if all $\rho_j = \rho_j^0$ (since $R_{V}^0 \equiv \sum_j \rho_j^0$). The second term represents the non-resonant absorption in the ground state as can be experimentally determined sufficiently before the resonance in the spectrum.

\begin{figure}[h]
\centering
\includegraphics[]{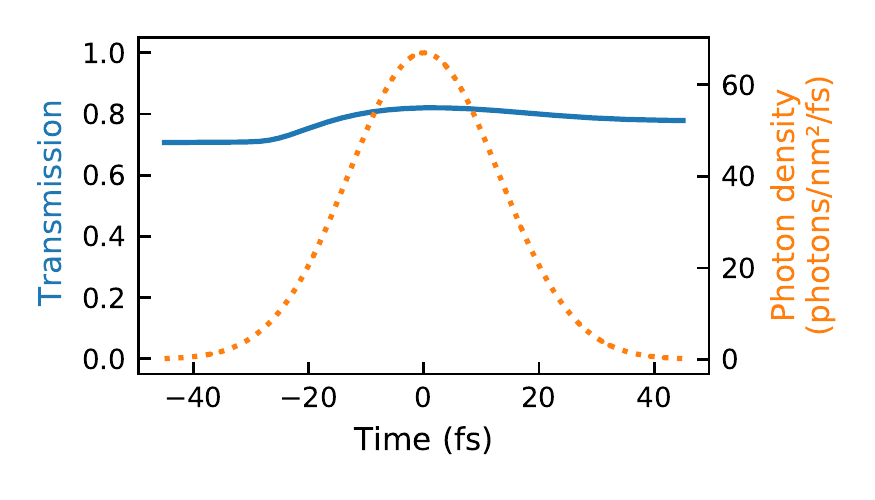}
\caption{
	\textbf{Instantaneous transmission} (including resonant and non-resonant absorption) over time for a pulse at 857.5\,eV with a pulse energy of 30\,$\mathrm{J/cm^2}$ (blue line, left axis), as well as the temporal profile of the incident photon density (orange dots, right axis). }
\label{fig: transmission}
\end{figure}

This treatment does not explicitly differentiate the non-resonant absorption from core energy levels other than the one treated by $R_C$. In the given example with photons resonant to the nickel 2$p$-absorption, the 3$s$ and 3$p$ core electrons only contribute to a minority of the non-resonant absorption events. In this model, we choose for simplicity to scale this contribution together with the non-resonant scattering from the valence electrons.

An exemplary incidence profile and the resulting transmission over time are shown in Figure \ref{fig: transmission}.

\subsubsection{Auger decay}
The model explicitly treats Auger decays that involve one core-hole and two electrons from the valence band. The rate at which an electron in density $\rho_j$ would decay via an Auger process is calculated as: 
\begin{equation*}
    P^\mathrm{Auger}_j = (M_C-R_C)\frac{\rho_j}{R_{V}^0}\frac{R_{V}}{R_{V}^0}  \frac{1}{\tau_{C}}
\end{equation*}
\begin{center}
\begin{tabular}{rl}
     $\tau^{C}$:& Core-hole lifetime \\
     $R_{V}$:& Total number of valence electrons\\
\end{tabular}
\end{center}
The first factor (in brackets) is the number of unoccupied core states, i.e. core-holes. The second factor describes the relative population of electrons at the energy $E_j$ and the third term is the relative population of the entire valence band to which the electron could transfer its energy. The latter two are normalized by the respective ground state population. The last term is the decay rate in the ground state, where $\tau_{C}$ represents the ground-state lifetime of a single core-hole.
Altogether, this describes Auger decays as interactions between two valence electrons, one emitted and one filling the $2p_{3/2}$ core-hole.
In reality, some fraction of Auger decays will emit electrons from the 3$s$ or 3$p$ core levels instead, followed by further Auger processes which emit electrons with the remaining energy of the original core-hole. These are not treated separately in our description, since the indirect decay is, on the one hand, a minority contribution and on the other hand, ultimately results in the same energy transfer to the valence band, albeit with a slightly longer time delay due to the intermediate steps.

\subsubsection{Free-electron scattering}
Inspired by earlier approaches to a simplified solution of the Boltzmann equation \cite{Bhatnagar1954BGKmodel}, we approximate the scattering rates of electrons in terms of characteristic time constants $\tau_{scatt}$ and $\tau_{th}$ for the free electrons and valence electrons, respectively. 

The lifetime of free electrons $\tau_{scatt}$ represents the inverse rate at which free electrons $R_{Free}$ scatter and decay to the valence system. While this parameter is ultimately empirical, it depicts a cascade of individual scattering events between electrons. In such a cascade, each free electron eventually transfers all its kinetic energy to the valence system.

\begin{equation*}
    P^\mathrm{scatt} = R_{Free} \frac{1}{\tau_{scatt}}
\end{equation*}
\begin{center}
\begin{tabular}{rl}
     $\tau_{scatt}$:& Free electron scattering time constant\\
     $R_{Free}$:& Number of free electrons\\
\end{tabular}
\end{center}

\subsubsection{Electron thermalization}
Similarly, $\tau_{th}$ characterizes the time with which the valence system approaches an internal thermal equilibrium.

\label{sec: el therm}
\begin{equation*}
    P^\mathrm{therm}_j = \left[ r_j(T, \mu) - \rho_j \right] \frac{1}{\tau_{th}}
\end{equation*}

\begin{center}
\begin{tabular}{rl}
     $\tau_{th}$:& Valence thermalization time constant\\
     $T$:& Equivalent electronic temperature\\
     $\mu$:& Chemical potential\\
\end{tabular}
\end{center}

To this end, the chemical potential and equivalent electronic temperature are calculated in each time-step based on the current internal energy  $U$ and number of valence electrons $R_{V}$.
The Fermi distribution for the calculated chemical potential and temperature then yields a momentary target electron distribution $r_j(T,\mu)$, which is approached with the electron thermalization constant $\tau_{th}$. 

\begin{align*}
    r_j(T, \mu) = m_j \frac{1}{e^{(E_j-\mu)/ k_B T}+1}\\
    U =\sum_j \rho_j E_j / \sum_j \rho_j\\
    R_{V} = \sum_j \rho_j
\end{align*}

\begin{center}
\begin{tabular}{rl}
     $U$:& Total energy of the valence system\\
     $R_V$:& Current total population of the valence system\\
     $r_j(T, \mu)$:& Electron density expected for\\
     &a fully thermalized valence system\\
     $k_{B}$:& Boltzmann constant\\
\end{tabular}
\end{center}

While the calculation of $r_j$ from $T$ and $\mu$ is straightforward, determining $T$ and $\mu$ from $U$ and $R_{V}$ is an inverse problem. This is solved by iterative optimization using the Levenberg-Marquardt method\footnote{First, the Levenberg-Marquardt root-finding algorithm is applied with a maximum of 400 iterations. If the required residual is still exceeded, the algorithm switches to least-squares optimization to refine the root-finding result. This combination proved a good compromise between computation speed and stability in regions with small gradients in the loss function.}. \\
Note that, because by definition both electron densities: the momentary $\rho_j$ and the thermalized goal distribution $r_j$ hold the same amount of electrons and internal energy, the sum over the valence band of all electron-distributing thermalization rates is always zero and the change in overall energy is also zero, i.e. $\sum_j P^\mathrm{therm}_j = 0$ and $\sum_j E_j P^\mathrm{therm}_j = 0$.

\begin{figure}[h]
\centering
\includegraphics[width=\linewidth]{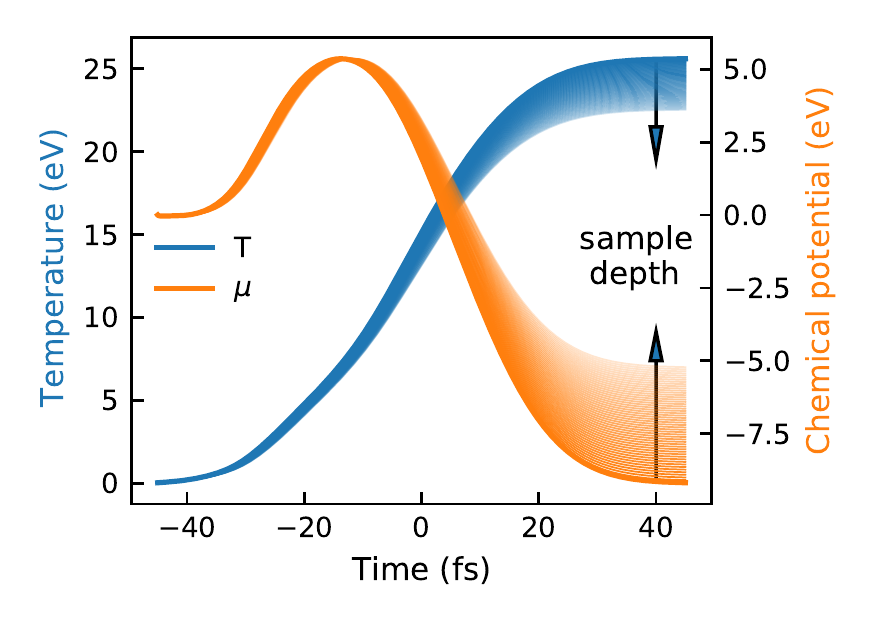}
\caption{
	\textbf{Temperature and chemical potential} over time for a pulse at 857.5\,eV with a pulse energy of 30\,$\mathrm{J/cm^2}$. The solid line represents the properties at the sample surface and the thin lines represent the deeper layers that are exposed to less X-ray fluence, as indicated by arrows and increasing transparency.}
\label{fig: fermi solver}
\end{figure}

Figure \ref{fig: fermi solver} shows how the temperature (given as an energy $k_B T$ in units of eV) and the chemical potential develop over time for a high fluence of 30\,$\mathrm{J/cm^2}$. Although reaching higher temperatures, these results are in agreement with studies treating the nickel valence system heated with optical lasers \cite{bevillon2014free, lin2008electron}.

\subsection{Differentials}
\label{sec: differentials}
From these process terms, we can assemble the time-differential of the various pools of electrons, photons, and energy. 
Because the simulation conserves the number of electrons in the sample, in the sum over all $j$, every process term describing electronic transitions appears equally often with positive and negative sign, representing a transition of electrons from one state to another. For example, the term for Auger decay appears twice with a negative sign in the valence electron differential, and each once with a positive sign in the differential for core- and free electrons. 

\subsubsection{Valence electrons}
\begin{equation*}
    \begin{split}
        \frac{d \rho_{j} }{dt} &= P^{res}_{i=j}  \\       
                            &- \sum_i P^\mathrm{non-res}_{i,j}\\
                            &- P^\mathrm{Auger}_j
                            - \frac{\sum_j \rho_j P^\mathrm{Auger}_j }{R_{V}} \\
                            &+ P^\mathrm{therm}_j \\
                            &+ \frac{h_j}{H_{V}} P^\mathrm{scatt} 
                            + P_{red} \left(\frac{-\rho_j }{R_{V}} + \frac{h_j}{H_{V}}\right)
    \end{split}
\end{equation*}

\begin{center}
\begin{tabular}{rl}
     $h_j$:& Number of valence holes, $h_j = m_j-\rho_j$\\
     $H_{V}$:& Total number of valence holes, $H_V = \sum_j h_j$\\
     $P_{red}$:& Electron re-distribution rate due to\\& scattering cascades from free electrons\\
\end{tabular}
\end{center}
The valence system interacts via all modeled processes, therefore we comment on the equation above line by line.
The resonant absorption rate $P^{res}_j$ changes the valence electron densities $\rho_j$ at all incident photon energies $E_i$ where $i=j$. 
Via non-resonant absorption, incident photon energies $E_i$ can remove electrons from $\rho_j$. 
The third line shows the primary and secondary Auger electrons. Note that in the sum over all energies $E_j$, terms 3 and 4 each remove one electron from the valence band for each Auger process occurring.
The thermalization term in line four drives electrons towards the thermal distribution based on the current internal energy and population of the valence band, without changing the total valence occupation, as discussed in section \ref{sec: el therm}.
The last line describes the effect of electron scattering.
The first term represents electrons from the free electron pool that are re-joining the valence system into a random unoccupied state $h_j$. 

The second term describes the electron redistribution inside the valence system in order to take up the energy released by the re-joining electron. 
This redistribution is calculated as a function of the rate of electronic scattering  $P^{scatt}$ and represents the effect of electronic scattering cascades. In such a cascade, a number of electrons is moved from occupied states $\rho_j$ to unoccupied states $h_j$.
The total rate of electrons redistributed in this time step through scattering $P_{red}$ is given by the ratio of the energy that needs to be taken up by the valence system to the energy that the valence system can additionally accommodate.
 \begin{equation*}
    P_{red}=\frac{S_{scatt}-S_{joining}}{U_h-U_e}
 \end{equation*}
The denominator of $P_{red}$ represents the energy that can maximally be redistributed to the valence system as the difference between the energy that could be contained in the unoccupied states
\begin{equation*}
      U_h = \frac{\sum_j h_j E_j}{H_V}
\end{equation*}
and the one already contained in the occupied states
\begin{equation*}
    U_{e} = \frac{\sum_j \rho_j E_j}{R_{V}}.
\end{equation*}
The numerator is given by the difference between the rates at which energy is released from the free electron energy pool
\begin{equation*}
    S_{scatt} = P^\mathrm{scatt} \frac{ E_{free}}{R_{free}} 
\end{equation*}
and the rate at which energy is gained in the valence system due to the formerly free electrons occupying random unoccupied valence states:
\begin{equation*}
    S_{joining} = \sum_j \frac{h_j}{H_{V}} P^\mathrm{scatt} E_j
\end{equation*}

\begin{figure}[h]
\centering
\includegraphics[width=\linewidth]{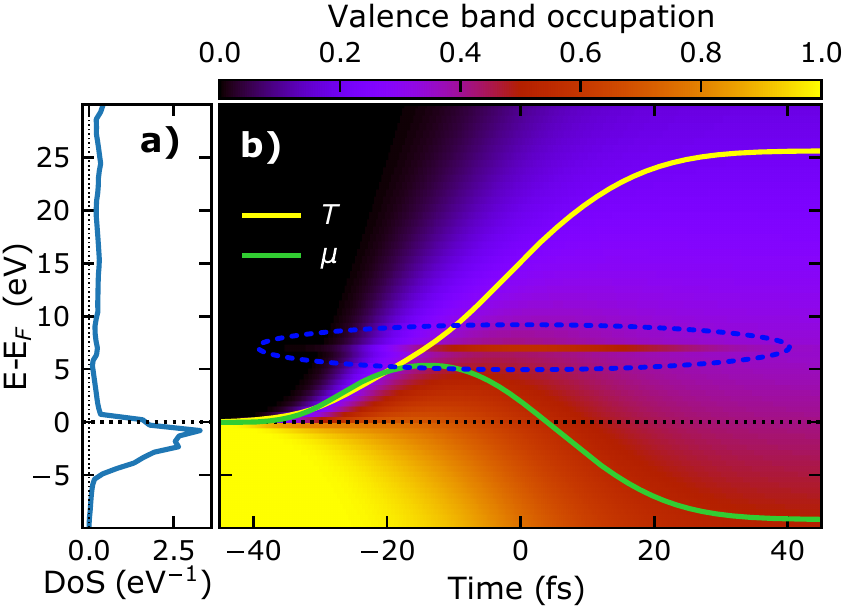}
\caption{
	\textbf{Density of states}. Panel a) shows the total DoS as used by the algorithm. Panel b) shows the relative population of the valence band over time at the sample surface for a pulse of 857.5\,eV with a pulse energy of 30\,$\mathrm{J/cm^2}$. The non-thermal saturation of the valence states (sometimes called spectral bleaching or hole burning) is clearly visible at the resonant energy 7\,eV above $E_F$ and highlighted with a blue ellipse. Apart from that, the shifting and broadening of the thermalizing valence distribution is apparent, together with the rising electronic temperature $T$ and shifting chemical potential $\mu$. The valence thermalization time is evident here in the visible lag of the thermalizing valence occupation behind the chemical potential.}
\label{fig: dos development}
\end{figure}

Figure \ref{fig: dos development} shows an example for the development of the valence system within a single voxel at the sample surface, exposed to a pulse of 30\,J/cm$^2$ fluence, resonant to states 7\,eV above $E_F$.

\subsubsection{Core electrons}
\begin{equation*}
    \frac{d R_{C} }{dt} =   - \sum_{i,j} P^{res}_{i,j}
                            + \sum_j P^\mathrm{Auger}_j
\end{equation*}

The population of core electrons is reduced (or increased, depending on the sign of $P^{res}_{i,j}$) by resonant transitions of all incident photon energies $E_i$ to states at all energies $\rho_j$ (although this contribution is only non-zero at $i=j$), and is increased by Auger decay from electrons of all energies $j$ in the valence system.
Note that the spontaneous emission channel is neglected in our model as it is designed for soft X-ray energies where Auger emission accounts for most core-hole decays (here specifically, 99.1\% of the nickel L$_3$ core-hole decays \cite{krause1979atomic, krause1979natural}).
In another concession to the specific experiment simulated here, we further neglect fast electrons leaving the sample, since the electron mean free path is much shorter than the sample thickness (about 1.3\,nm \cite{powell2020NIST} compared to a 20\,nm thick sample).
While a loss process for free electrons would be trivial to implement, the total number of electrons in the system being strictly constant over time is a valuable indicator for the self-consistency of the calculation.

\subsubsection{Free electrons}
\begin{equation*}
    \frac{d R_{Free} }{dt} =  \sum_{i,j} P^\mathrm{non-res}_{i,j}
                            + \sum_j P^\mathrm{Auger}_j
                            - P^\mathrm{scatt} 
\end{equation*}

Unbound or free electrons are generated by non-resonant absorption from all incident photon energies $E_i$ as well as Auger-decays from all energies in the valence band. The population is reduced by the free electron scattering rate $P^\mathrm{scatt}$.

\subsubsection{Photon absorption and emission}
\begin{equation*}
    \frac{d N_{i} }{dz} =  - P^{res}_{i=j}
                           - \sum_{j} P^\mathrm{non-res}_{i,j}
\end{equation*}
The number of photons is reduced or increased by resonant interaction and reduced by non-resonant absorption.
Note that this is a purely spatial differential that depicts the instantaneous transmission of a certain number of photons through the entire sample in each time-step.

\subsubsection{Energy of free electrons}
\begin{equation*}
    \begin{split}
        \frac{d E_{Free} }{dt} &=  \sum_{i,j} P^\mathrm{non-res}_{i,j}(E_F+E_i-E_j)\\
                            &+ \left(\frac{\sum_j E_j (\rho_j - P^\mathrm{Auger}_j)}{\sum_j (\rho_j - P^\mathrm{Auger}_j)}+E_F\right)\sum_j P^\mathrm{Auger}_j \\
                            &- P^\mathrm{scatt} \frac{E_{Free}}{R_{Free}}
    \end{split}
\end{equation*}

\begin{center}
\begin{tabular}{rl}
     $E_F$:& Energy of the Fermi level\\
\end{tabular}
\end{center}
Because photons can "kick out" electrons from arbitrary states in the valence system, it becomes necessary to track the total kinetic energy of the free electrons, even though the distribution of energy among these electrons is not tracked.
The rate of energy transfer to the free electron bath is described as the sum of non-resonant absorption and Auger decays, each multiplied by their respective energies (first and second line, respectively). The energy of the secondary Auger electrons is calculated as the average energy of all electrons other than the primary Auger electrons, as those drop to the core level. 
Finally, each electron that leaves $R_{Free}$ reduces the energy of the bath by the average energy, which is $\frac{E_{Free}}{R_{Free}}$.

\begin{figure}[h]
\centering
\includegraphics[width=\linewidth]{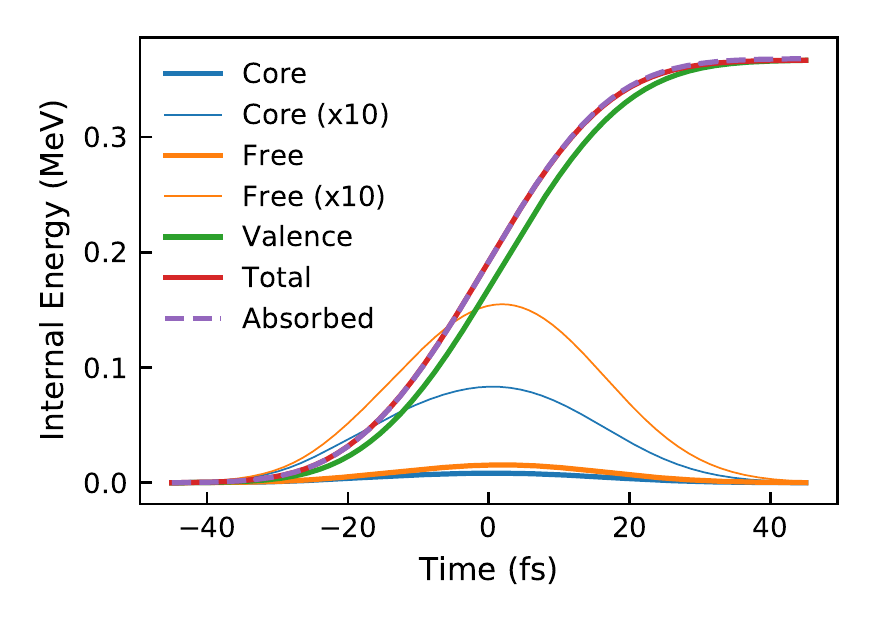}
\caption{
	\textbf{Energy in the sample system} over time for a pulse at 857.5\,eV, integrated over the full 20\,nm thickness of the sample and a 1\,nm$^2$ area illuminated with a fluence of 30\,$\mathrm{J/cm^2}$. The absorbed energy is calculated from the difference between incident and transmitted photons, while the total energy is a sum of the energy held in the electronic sub-systems of core-holes, free electrons, and valence excitation. 
    Due to the fast process rates in comparison to the pulse duration, the energy held in core excitations and free electrons remains small, which is why a 10 times scaled curve is also shown.}
\label{fig: energy conservation}
\end{figure}

Explicitly tracking this energy also enables us to demonstrate the conservation of energy within the simulation; since there is no channel that allows energy to leave the sample, the energy held in the electronic sub-systems matches that of the absorbed photons at all times.  Figure \ref{fig: energy conservation} shows the internal energy of the electronic subsystems over time, integrated over an area of 1\,nm$^2$ and the full 20\,nm thickness of the sample.
The comparison of the energy in various sub-systems demonstrates how quickly the energy of the photon pulse is transferred to valence excitations.
Furthermore, observing the energy conservation has proven to be an invaluable tool to select sufficiently fine binning in time, space and energy, as it is particularly sensitive to the accumulation of numerical errors. For example, the calculation for the homogeneous illumination with 30\,$\mathrm{J/cm^2}$ shows a cumulative error in energy of 0.35\%. This is in contrast to the electron conservation, which is strictly kept to machine precision level due to the symmetric way the process terms are arranged to form the time differentials.

\subsection{Parameters}
\label{sec: parameters}
There are four categories of parameters: 
First, resolution parameters for the number of steps in time, space, and energy are chosen as a compromise between calculation time and numerical error; 
second, experimental parameters such as the pulse duration, peak fluence, and bandwidth of the interacting photons reflect experimental conditions; 
third, ground-state properties such as the atomic density and number of states and electrons, as well as Auger-decay limited core-hole lifetime are drawn from published literature, while the resonant and non-resonant absorption lengths are derived from the ground-state spectrum as described below.
Fourth, the model-inherent phenomenological parameters are the valence thermalization time $\tau_{th}$ and electron scattering time $\tau_{scatt}$, which are varied to achieve the best match to the experimental results.
A list of the relevant parameters and the chosen values for the present calculations, selected to match the experiment presented in \cite{engel2022XFEL}, is shown in the appendix Table \ref{tab: parameters}.

\begin{figure}[h]
\centering
\includegraphics[width=\linewidth]{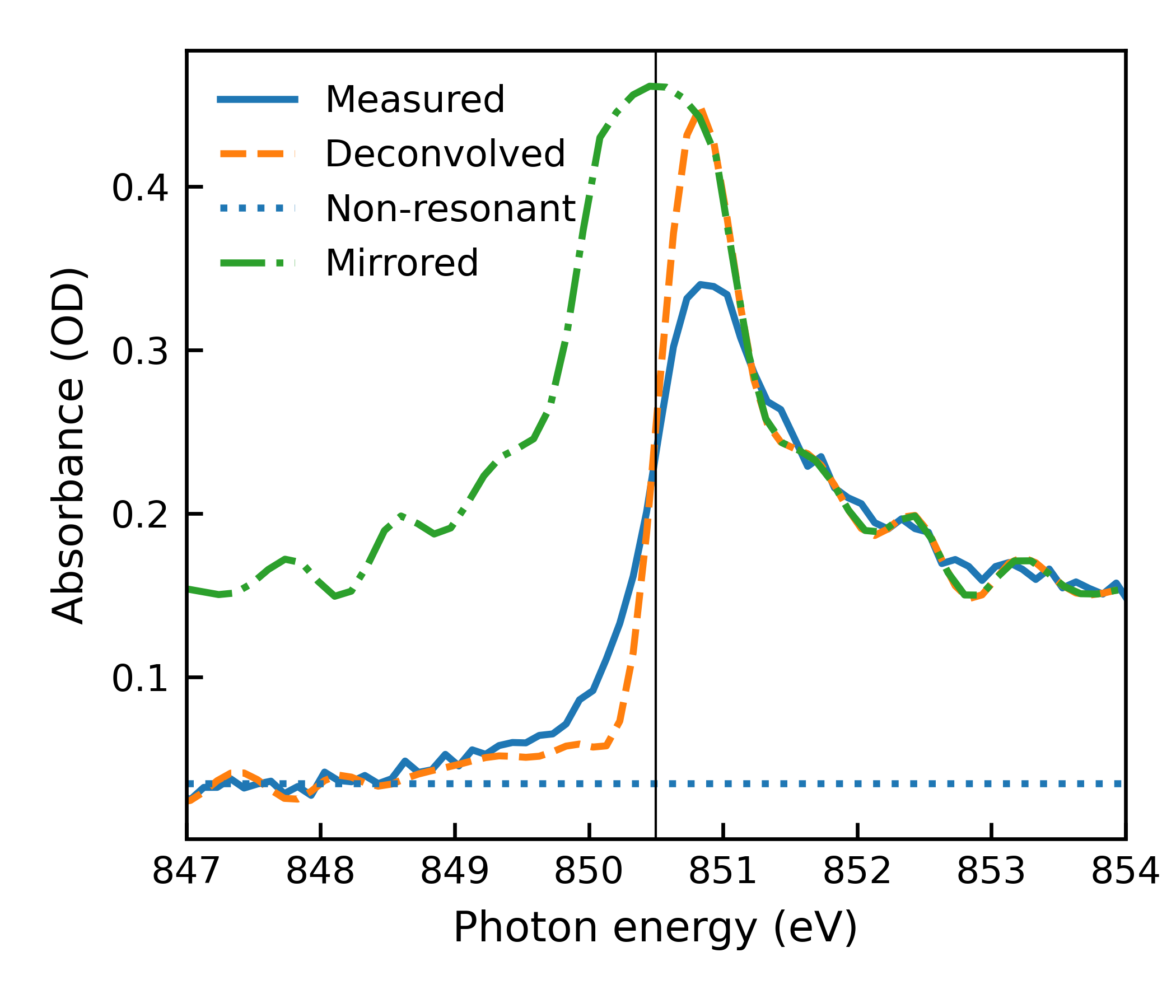}
\caption{
	\textbf{Derivation of absorption lengths as input parameters}, reconstructed from a measured ground-state spectrum (blue line). The non-resonant absorption level (blue dots) was determined from the pre-edge region. The measured resonant absorption length was deconvolved with the experimental resolution (orange dashes) and mirrored around the rising edge to retrieve a symmetric resonant absorption length around the resonance (green dot-dashed line). See main text for details.}
\label{fig: input spectrum}
\end{figure}

The parameterization of several ground-state properties deserves further comment.
Firstly, the DoS (motivated at the start of this section) was subdivided into bins $m_j$ of varying size, favoring a fine resolution for the bound states. 
The size of the energy bin that is resonantly coupled to the core level by the incident photons is chosen such that it represents the interaction bandwidth of the photons.
The interaction bandwidth is calculated as the convolution of the bandwidth of incident photons (i.e. the resolution of the experiment) and the natural line width of the core excitation. We further account for the final state broadening of excitations into less tightly bound states by enlarging the interaction bandwidth by 0.1\,eV per eV above the Fermi level.
Furthermore, the resonant and non-resonant absorption lengths are derived from the ground-state spectrum. We treat the non-resonant absorption length as constant, i.e. independent of photon energy, and derive it from the pre-edge absorption level.
The transition matrix element of a core-valence transition exhibits a resonant enhancement close to the absorption edge, which translates into an energy dependence of the resonant absorption length.
Above the Fermi level, where the DoS is unoccupied in the ground state, the resonant absorption length is encoded in the ground-state absorption spectrum.
As the transition matrix element from the core level to states below the Fermi level is experimentally not straightforward to access, we use the approximation that the energy-dependence of the transition matrix element is symmetric around the Fermi energy. 
To derive the resonant absorption length, the non-resonant absorption level is subtracted from the spectrum and line broadening is accounted for by deconvolution with a pseudo-Voigt-profile of 50\% Gaussian and Lorentzian share and a width of 640\,meV Full-Width Half-Maximum (FWHM), representing 420\,meV broadening from the experimental resolution and 480\,meV from the core-hole lifetime \cite{krause1979natural}.
The deconvolved resonant absorption spectrum above the Fermi level is then mirrored around the Fermi energy and the discontinuity within 320\,meV around it is reconstructed with cubic interpolation. This results in the mirrored resonant absorption spectrum shown in Figure \ref{fig: input spectrum}, which is used as the resonant absorption length parameter $\lambda_i^{res}$.
Note that the results of simulated spectra were finally re-convolved with the same pseudo-Voigt-profile to simulate the same experiment.

\section{\label{sec: discussion} Discussion}

\subsection{Parameter Study}

\begin{figure}
    \centering
    \includegraphics[width = \linewidth]{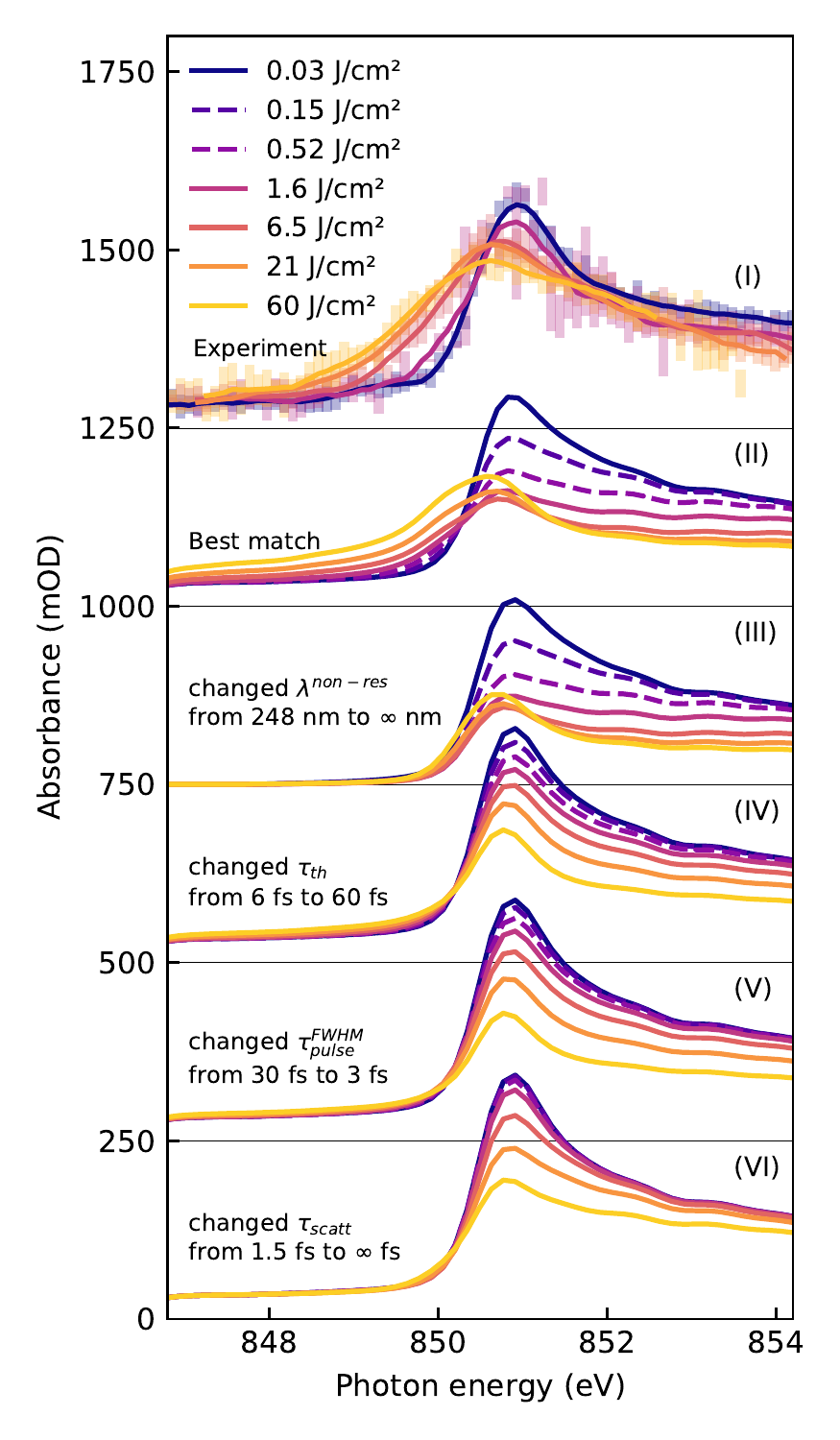}
    \caption{\textbf{Fluence-dependent Ni $L_3$-edge spectra, simulated with different parameters and compared to the measurements presented in \cite{engel2022XFEL}} For the spectra (III) to (VI), one parameter was varied with respect to the best batch (II). Each set of spectra is offset by another 250\,mOD as indicated by the horizontal lines.
    The error bars of the experimental data represent the 95\% confidence intervals for each bin of 102\,meV width; the solid lines represent smoothed spectra using a Savitzky-Golay filter using windows of 21 bins and 4th-order polynomials. The average fluence of events contributing to each spectrum is given in the legend. Dashed simulated spectra do not have a corresponding measurement.
    }
    \label{fig: spectra}
\end{figure}

In Figure \ref{fig: spectra} we show the measured non-linear X-ray absorption spectra labeled (I) presented and discussed in \cite{engel2022XFEL} together with sets of simulated spectra computed for different sets of parameters. 

The first set of simulated spectra (also presented in \cite{engel2022XFEL}) represents the best match with the experimental conditions and the parameters shown in {table \ref{tab: parameters}} and is labeled (II), while the consecutive sets, labeled (III) to (VI), demonstrate how the results change when individual parameters are modified. 
We present this set of simulations to showcase how our model may be used to understand the relation between the non-linear changes and various parameters. 
To fit the experimental results, however, only the parameters $\tau_{th}$ and $\tau_{scatt}$ are treated as unknowns, while all other parameters are known experimental or ground-state parameters.

In the best matching simulation (II), the experimental observations of a red-shifted rising edge, increased pre-edge absorption, as well as reduced absorption at and beyond the absorption peak, are reproduced. 
However, the saturation effect at the resonance is over-estimated and the lack of measured saturation around 852\,eV cannot be explained by our model. See \cite{engel2022XFEL} for a more detailed discussion of these deviations.

The next simulation (III) was performed without non-resonant absorption. While this eliminates the pre-edge absorption rise, some shift of the absorption onset is still predicted within the original peak, while the spectra above the resonance onset behave similarly to the best-matching simulation.

With simulations (IV) and (V), we demonstrate the effect of prolonging the thermalization time or shortening the pulse duration by a factor of ten, respectively. Both have the similar effect of strongly reducing the peak shift and a moderate decrease of saturation beyond the resonance.

Finally, we show non-linear spectra (VI) where the free electron scattering process was eliminated ($\tau_{scatt}=\infty$). This prevents the majority of the excitation energy from entering the valence system and thus drastically reduces valence heating.
This causes the pre-edge absorption rise and rising edge shift to vanish nearly entirely, and absorption-decrease due to saturation at and above the resonance is also strongly reduced. 
These differences underscore the importance of electronic scattering cascades to these phenomena. 

While an extensive study of the correlations between the various model parameters and the observed effects is beyond the scope of this work, this brief parameter study allows an interpretation of how red-shift and pre-edge absorption rise occur:
Let us consider a case where the incident photon energy is slightly below the absorption edge. Initially, only non-resonant absorption transfers energy to the sample, specifically by creating free electrons in form of photo-electrons. 
This energy is then transferred to the valence band due to electronic scattering cascades, where it causes many small, non-thermal excitations. These are homogeneously distributed over all valence energies and thus only lead to a small increase in pre-edge absorption (see the simulation with slow thermalization).
However, this distribution of excitations develops towards the shape of a Fermi distribution over time, specifically with the thermalization time constant. 
If the pulses are long in comparison to the thermalization time, a significant number of states below the Fermi level becomes available for core-valence transitions during the pulse duration. This causes the observed increase in pre-edge absorption.
Once the first empty valence states become available at the current photon energy, 
resonant absorption begins to occur in addition to the non-resonant absorption.
The additional resonant absorption leads to more free electrons from Auger-decay, which in the same way as the photo-electrons further contribute to secondary electron scattering and thermalization. Because thermalization mostly creates free states just below the Fermi level, this process of self-enhancing rise of overall absorption manifests in a fluence-dependent shift of the rising edge to lower energies.

\subsection{Limitations and Potential}
As demonstrated in the parameter study above, this kind of analysis can provide a straightforward interpretation of how non-linear changes to the absorption spectrum emerge, and what each change says about the development of the electronic populations.

While the model already demonstrates good agreement with experimental results \cite{engel2022XFEL}, below we discuss the limitations of our approach and possibilities for expanding or refining it.

In the model as described here, the DoS is not only assumed constant but also does not differentiate between spin-up and spin-down states, although we are treating a magnetic material. 
Splitting up the DoS in spin-up and spin-down states would allow for the inclusion of angular momentum conservation and transfer in the various scattering rates. 

For thin samples where the electron mean free path becomes similar to the layer thickness, a loss process for free electrons should be introduced to account for electrons leaving the sample.

Fluorescent decay should be accounted for when moving to harder X-rays \cite{krause1979atomic}. While introducing the decay channel itself would be simple, accurately accounting for reabsorption may be less straightforward, since the model only propagates light in one direction.

Furthermore, the thermalization time $\tau_{th}$ is used in this work as a global fitting parameter, although electron thermalization times have been suggested to depend on electronic temperature \cite{mueller2013relaxation}.
Since the electronic temperature and target distribution are calculated every time-step, an arbitrary dependence could be easily introduced, albeit with the necessity of additional fitting parameters.

The DoS is dominated by the crystal lattice, which is typically stable on the sub-100\,fs timescale. However, recent Time-Dependent Density Functional Theory (TDDFT) calculations show that electronic processes (i.e. sub 100\,fs) can also lead to modifications of the DoS via spin-orbit coupling modeled by introducing an onsite Hubbard correlation $U$ to the mean-field Hamiltonian \cite{Lojewski2022XAS}.
Since the rate model approach is generally not suited to calculate the DoS, incorporating modifications to the DoS would require a close interplay of the rate model with (TD)DFT calculations, leading to an ultimately much more complex approach reminiscent of models developed for the study of radiation-induced damage mechanisms \cite{medvedev2018various, lipp2022density}.

Furthermore, we derive the interaction bandwidth (the valence energy range to which core states can be resonantly coupled by incident photons) as a convolution of instrumental resolution and the lifetime of the core excitation, i.e. the Auger lifetime. The final state lifetime broadening of excitations into continuum states is described as a continuous broadening of 0.1\,eV per eV above the Fermi energy \cite{stohr1992nexafs}.
It is however reasonable to expect that the final state lifetime is further shortened at higher fluences, due to increased rates of both electronic scattering and stimulated emission; the latter would be particularly relevant at the resonance peak. Such further broadening would cause more valence states to be available for resonant interaction and reduce the observed saturation effect while increasing the number of core-holes that may be created by high fluences\footnote{When considering only resonant absorption, the core level, and resonant valence states constitute a classic two-level system. In the extreme fluence limit, the relative populations of such a system are given by the state degeneracy ratio, i.e. the ratio between the number of involved core and valence states.}.
Accounting for a fluence-dependent broadening of the interaction bandwidth in a refined rate model might help to remedy the overestimation of the saturation effect at the absorption peak in the presented calculations.

Another candidate for further refinement is an energy dependence of the electronic scattering rates. The presented model uses fixed rates for thermalization and scattering cascades, which both act on all valence states indiscriminately. This description would be especially inadequate when applied without modifications to a bandgap material.
An advanced model could describe both the thermalization rate of the valence band and the energy of excitations from scattering cascades in an energy-resolved manner\footnote{It might be non-trivial to normalize this energy dependence such that the electronic scattering processes do not violate the conservation of energy.}.

While such refinements may seem attractive, a core strength of the rate model approach is its relative simplicity and computational tractability, as well as the use of known ground-state parameters, which supports a straightforward physical interpretation. It is ultimately a mostly classical, phenomenological model which offers a complementary approach to ab-initio calculations.
Every added complexity should therefore be weighed against its relevance, as a simpler model facilitates a meaningful interpretation and avoids introducing ambiguity in the results due to correlations between redundant input parameters.

Since the model operates on widely applicable principles, we expect that it may be applied to a wide range of materials with some predictive power, while the limitations described above apply. 
The results will especially deviate from observations wherever multi-particle effects, such as electron correlations or quasiparticles become relevant.

As presented, the model enables an understanding of the electronic population history under strong X-ray fluences and characterization of the resulting non-linear absorption near a core resonance.
Non-linear absorption studies like the one analyzed here \cite{engel2022XFEL} allow one to characterize the excitation dynamics of the electronic system under study. Furthermore, consideration of these dynamics is, due to the extreme fluences required, particularly relevant for methods in the emerging field of non-linear X-ray spectroscopy \cite{mukamel2005multiple, Glover2012xray, beye2013stimulated, weninger2013stimulated, Shwartz2014XSHG, Tamasaku2014TPA_competing, Bencivenga2015FWMTG, schreck2015implications, Lam2018HXSHGinterfacial, Tamasaku2018NL_TPA, higley2019femtosecond,Beye2019musix, rottke2021probing, Rouxel2021HXTG, Bencivenga2021FWMtwo_color, Wirok_Stoletow2022TPA_Ge, Boemer2021XoptWMprobes, Serrat2021locFWM}.

\section{\label{sec: conclusion} Conclusion}
In this work, we propose a model of differential rate equations describing the various excitation and decay processes that connect core- and valence electronic states to quantitatively describe the non-linear changes to X-ray absorption around a core resonance that occur when employing increasingly high FEL pulse energies. We present the framework of the model in detail, applying it to the case of recently measured non-linear absorption spectra of nickel films at the $L_3$ edge, recorded with monochromatic X-rays \cite{engel2022XFEL}.

We demonstrate how the rate model reaches good agreement with the experimental results while disentangling the contributions of the relevant physical processes: resonant and non-resonant absorption, Auger decay, electron thermalization, and electronic scattering cascades.
Our rate model contains numerous simplifications in order to approximate the result of complex interactions between many particles in terms of average transition rates.
Nevertheless, the resulting description enables a quantitative understanding of the evolution of the system and the processes responsible for spectral changes.
In the presented calculations, the most relevant effect is valence system heating due to secondary electron cascades from free electrons.
Our results allow quantifying the absorbance changes caused by fundamental electron population dynamics, which are crucial to disentangle from collective quantum effects studied with currently evolving non-linear X-ray spectroscopy techniques.

\begin{acknowledgments}
Funding by the Helmholtz Association (grant VH-NG-1105) is gratefully acknowledged. This research was supported in part through the Maxwell computational resources operated at Deutsches Elektronen-Synchrotron DESY, Hamburg, Germany\\
\textbf{Author Contributions}: 
R.Y.E. and M.B. developed the model.  R.Y.E. wrote the manuscript with contributions from all authors.
\end{acknowledgments}

\clearpage
\appendix

\counterwithin{figure}{section}

\section{Run-Time and Code Availability}
Using adequate multiprocessing on a single node of a computing cluster allows computation of non-linear spectra such as presented here within many minutes to several hours, depending on the required number of steps in space, time, and energy. The simulation code is publicly available \cite{githubXNLdynCode}.

\hspace{2cm}

\section{Parameter table}
\begin{scriptsize}
\begin{longtable*}{|l|l|l|l|l|} 
    \hline
    \textbf{Symbol} & \textbf{Code} & \textbf{Description} & \textbf{Unit}& \textbf{Value}\\
    \hline
    $N_z$           & \texttt{Nsteps\_z}  & Steps in sample depth       & -&50\\
    $N_{E_j}$       & \texttt{N\_j}       & Steps in energies considered in valence system  & - & 90 \\
    $N_{E_i}$       & \texttt{N\_points\_E} & Number of photon energies / points in the spectrum   & - & 69 \\
    -              &\begin{tabular}{l} \texttt{N\_local\_fluences}\\ \texttt{\_to\_calculate}  \end{tabular} & Number of fixed fluences that are directly simulated& - & 30  \\
    -           & \texttt{N\_pulse\_energies}  & Number of final pulse energies with a Gaussian spot profile& -&20  \\
    -           & \texttt{Nsteps\_r}  & Number of steps in the radial integration of the Gaussian spot& -& 100  \\
    $dt_{min}$      & \texttt{timestep\_min}         & Minimum allowed time-step          & fs & 0.15\\
    
    -               & \texttt{Energy\_axis\_max}& Maximum energy in the valence system & eV &800\\
    -               & \texttt{Energy\_axis\_fine\_until}& Finer sampling  for energies lower than this  & eV & 30\\
    -               & \texttt{Energy\_axis\_min}& Valence band origin   & eV &-10\\

    -               & \texttt{DoS\_band\_origin}& Energy minimum from where to use the loaded DOS  & eV &-10\\
    -               & \texttt{DoS\_band\_dd\_end}& Energy maximum from where to use the loaded DOS  &  eV &30 \\

    \hline
    $\sigma_{tj}$   & \texttt{tdur\_sig}   & Rms pulse duration of photons &  fs & 13\\
    $E_i$           & \texttt{E\_i}  & Photon energy of incident photons &  eV &848-856\\
    $I_0$           & \texttt{I\_0}  & Number of photons irradiated      & photons nm$^{-2}$ & $0-1.4\text{e}4$\\
    $N_{E_i}$       & \texttt{N\_photens}   & Number of different photon energies irradiated   & - & 1\\
    $T_0$           & \texttt{temperature}   & Initial sample temperature  & K & 300\\
     $\sigma_{BW}$  & \texttt{interaction\_bandwidth}  & Bandwidth of resonant interaction at $E_j$ &  eV & 0.638\\
    \hline
    $\tau_{scatt}$  & \texttt{tau\_scattering} & Scattering time of free electrons  & fs & 1.5\\
    $\tau_{th}$ & \texttt{tau\_th} & Thermalization time of non-thermal valence states & fs & 6\\  
    \hline
    -	     	    & \texttt{DoS\_shapefile}  & Filename of the total DOS from DFT-calculation & - & from \cite{Persson2016nickel}\\   
    $Z$             & \texttt{Z}         & Total sample thickness           & nm & 20\\
    $\rho$          & \texttt{atomic\_density} & Atomic density  & atoms nm$^{-3}$ & 91.4\\
    $R_V^0$	     	    & \texttt{valence\_GS\_occupation}  & Valence electrons per atom in the ground state& states atom$^{-1}$ & 10\\
    $M_C$	     	& \texttt{core\_states}  & Core electrons/states per atom in the L$_3$ core level& states atom$^{-1}$ &4\\
    $E_f$           & \texttt{E\_f}     & Fermi level; used as zero for energy axes of $E_j$ and $E_i$   & eV&850.5\\
    $\tau_{C}$     & \texttt{tau\_core\_hole} & core-hole lifetime, from \cite{krause1979natural} & fs  & 1.4 \\
    $\lambda^{non-res}$  & \texttt{lambd\_nonres} & Absorption length due to non-resonant absorption  & nm &248\\
    $\lambda_{Ej}^{res}$  & \texttt{lambd\_res\_Ej} & Absorption length due to resonant absorption  & nm & 20-83 \\

    \hline
    \caption{Parameters for the presented simulation results. The first block lists parameters that define the resolution of the simulation, the second block shows experimental conditions, the third phenomenological fitting parameters and the last block contains physical ground-state properties of the sample.}
    \label{tab: parameters}
\end{longtable*}
\end{scriptsize}




\newpage
\bibliography{bib-nl-xas}

\end{document}